\documentclass[aps,pre,showpacs,floatfix,twocolumn]{revtex4}
\usepackage{graphicx}

\newcommand{\la}{\langle}
\newcommand{\ra}{\rangle}

\newcommand{\lbsq}{\overline{\ell^2}}
\newcommand{\bea}{\begin{eqnarray}}
\newcommand{\eea}{\end{eqnarray}}
\newcommand{\Nc}{N_{\mbox{cov}}}

\begin{document} 

\title{Scaling Properties of Random Walks on Small-World Networks} 
\author{E. Almaas}
\email{Almaas.1@nd.edu}
\affiliation{Department of Physics, University of Notre Dame, Notre Dame,
Indiana 46556} 
\author{R. V. Kulkarni\cite{new}}
\email{rahul@research.nj.nec.com}
\affiliation{Department of Physics, University of California, Davis, California 95616}

\author{D. Stroud} 
\email{stroud@mps.ohio-state.edu}
\affiliation{Department of Physics, The Ohio State University, Columbus, 
Ohio 43210} 

\date{\today} 
\begin{abstract} 
Using both numerical simulations and scaling arguments, we study the
behavior of a random walker on a one-dimensional small-world network.
For the properties we study, we find that the random walk obeys a
characteristic scaling form.  These properties include the average
number of distinct sites visited by the random walker, the mean-square
displacement of the walker, and the distribution of first-return
times.  The scaling form has three characteristic time regimes.  At
short times, the walker does not see the small-world shortcuts and
effectively probes an ordinary Euclidean network in $d$-dimensions.
At intermediate times, the properties of the walker shows scaling
behavior characteristic of an infinite small-world network.  Finally,
at long times, the finite size of the network becomes important, and
many of the properties of the walker saturate.  We propose general
analytical forms for the scaling properties in all three regimes, and
show that these analytical forms are consistent with our numerical
simulations.
\end{abstract} 
 
\maketitle 

\section{Introduction}

The topological properties of real-world networks have been studied
extensively.  But an even more intriguing task, and a natural
extension of these studies, is to understand how the network structure
affects dynamics on the networks.\cite{strogatz01} Most people have
had the unfortunate experience of catching the flu (an example of
disease spreading) or picking up a burning hot plate (an example of
neural signal transmission).  These phenomena are all examples of
dynamics on the special kind of real-world networks that have been
found to display ``small-world'' properties.  Of these phenomena, the
greatest attention thus far has been given to the study of disease
spreading [see, e.\
g. \cite{albert01,kuperman01,newman02,satorras01}].  It has even been
suggested that the web of sexual contacts has a small-world structure
\cite{deszo01,liljeros01}.  Other dynamical models that have been
studied on complex networks include the Hodgkin-Huxley model
\cite{lago00}, Boolean dynamics \cite{reka}, and the generic
synchronization of oscillators \cite{pecora}.  Extensive reviews can
be found in Refs.\ \cite{albert01,dorogovtsev01,strogatz01}.

In this paper, we will present results for a random walk on a
small-world network (SWN, defined below).  Such random walks may have
several applications to real systems.  For example, Scala {\it et
al}\cite{scala01} have argued that the conformation space of a lattice
polymer has a small-world topology, and hence, that diffusion and
random walks on such small-world networks might give insight into
relaxation processes such as protein folding.

Much is known about random walks on both regular and random networks
(see, e.\ g., \cite{barber70,hughes}).  In addition, there have been
several recent studies of random walks on SWN's
\cite{monasson99,jespersen00,jasch01,lahtinen01,pandit01,scala01,octavio,lahtinen02}.
For example, Jasch and Blumen\cite{jasch01} and Lahtinen {\it et. al}
\cite{lahtinen01} have studied the average number of distinct sites
visited by a walker, and the probability that the walker is at the
origin after $n$ steps in the limit that the network size $L
\rightarrow \infty$ and, hence, the number of shortcuts $x \rightarrow
\infty$.  In the present work, we present results for the scaling
behavior of such quantities as the mean-square displacement, the mean
number of distinct sites covered during a random walk, the
first-return time abd clarify the procedure for obtaining a scaling
collapse for random walks on SWN's.

\section{The random walk}

In previous work \cite{kulkarni,almaas02_1}, we developed a simple
probability function approach for the topological properties of
complex networks generated according to the small-world
model\cite{watts98,newman99}.  In this approach, we start from a
one-dimensional regular network with periodic boundary conditions and
$L = 2 N$ nodes, each node being connected to its $2 k$ nearest
neighbors. Hence, the ``degree'' of each node is $2 k$.  Next, we add
shortcut ends to each node according to a given degree distribution
(${\cal D}_q$), by following the prescription of Refs.\
\cite{molloy,newman01}. In the present work we use a modified approach
\cite{almaas02_1} by using the following degree distribution in one
dimension (with $k=1$): ${\cal D}_q = (1-p) {\cal P}_{(q)}(p) + p~
{\cal P}_{(q-1)}(p)$, where ${\cal P}_q (\lambda) = \exp(-\lambda) ~
\lambda^q / q!$ is the Poisson distribution.  We then select pairs of
shortcut ends at random and connect them to each other, thus creating
a shortcut. This network-generating procedure (with the above ${\cal
D}_q$) is equivalent to that outlined by Newman and Watts
\cite{newman99}.  The quantity $k p$ is the probability that a given
site has a shortcut.  On average, there will be $x = k p L$ shortcuts
in the network.

We carry out the random walk on a lattice (Polya walk) as follows: (i)
There is only one walker on the small-world network at a time.  (ii)
The random walker is injected onto a randomly chosen site on the
small-world network, a new site for each walker.  We will call this
site the ``origin'' of the walk. (iii) At each discrete time step,
$t$, the walker will jump to a randomly chosen nearest neighbor of its
current site $m$ with probability $1/k(m)$.  Here, $k(m)$ is the
number of nearest neighbors of site $m$, {\it i.e.} the degree of node
$m$. (iv) The random walker is allowed to wander the network for a
time longer than the ``saturation time'' for the quantity studied,
{\it i.e.} the time when that quantity approaches its limiting
behavior.  (v) We average over different random walkers and
realizations of the small-world lattice until the results converge.

\section{Scaling Behavior}

\subsection{General Form}

In earlier work \cite{kulkarni,almaas02_1}, we demonstrated that the
basic probability distribution $P(m|n;L,p)$ (the probability that two
sites separated by $n$ hops before the introduction of shortcuts has a
minimal separation of $m$ hops when shortcuts are included) scales
with $x$ in the limit of $p \ll 1$, for all choices of $x$. As a
consequence, topological quantities derived from $P(m|n;L,p)$ displays
a scaling with $x = p L$ in the same limit.

We will first state our main result for the scaling of the random walk
in general, and then consider specific examples. Let $O(p,L,t)$ be
some measurable quantity for the random walk on an SWN, which
saturates to a finite value $O_{sat}$ as $t \rightarrow \infty$.  As
specific examples, we will discuss the mean-square displacement and
the average number of sites covered by the random walker in time
$t$. We propose that $O(p,L,t)$ satisfies the scaling law:
\begin{equation}
	O(p,L,t) = O_{sat} ~ {\cal F} (p^2 t,p L). 
	\label{eq:genscale}
\end{equation}
We have numerically confirmed this scaling Ansatz, as described in the
rest of this paper.  According to Eq.\ (\ref{eq:genscale}), a scaling
collapse is observed if one plots the quantity of interest for various
choices of the variables $(p,L)$, holding $x = p L = \mbox{const.}$,
for {\em any} choice of $x$.  This behavior is in contrast to previous
statements that a scaling collapse cannot be expected when $x \gg
100$ \cite{lahtinen01}.  Note also that the scaling collapse is seen
only for {\em fixed} values of $x = p L$ and that a prerequisite for
scaling is that $p \ll 1$. Previous workers \cite{jasch01,lahtinen01}
have attempted to show scaling collapse for fixed values of $L$ (or
$p$) while varying $p$ (or $L$), and did not obtain a perfect scaling
collapse.  By contrast, our present results display a perfect scaling
collapse for fixed values of $x$. In the following, we present our
numerical results for two quantities: the mean number of distinct
sites covered, $\Nc$, and the mean square displacement, $\la r^2 \ra$.

\subsection{Average Number of Distinct Visited Sites}

\begin{figure}[tb]
\centerline{\includegraphics[height=6.5cm]{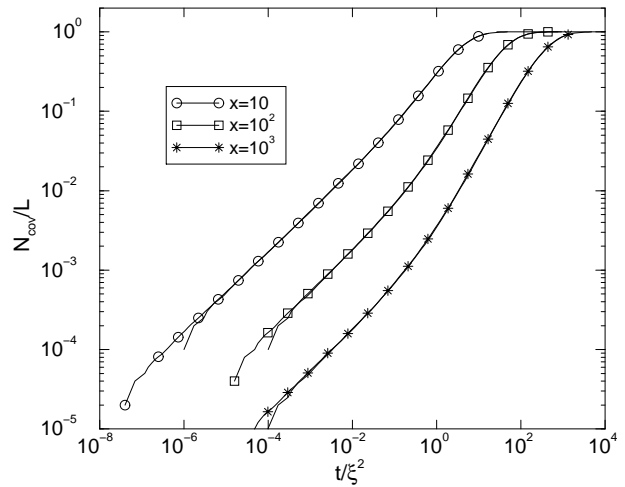}}
\caption{This figure shows the scaling collapse for $N_{\mbox{cov}}$,
the average number of distinct sites visited by the random walker, for
$x = p L = 10$ (circles, system sizes $L = 10^4$ and $L = 10^5$);
$x=100$ (squares, $L=10^4$ and $2.5\times 10^4$; and $x=1000$
(asterisks, $L=10^5$ and $L=2.5\times 10^5$). The scaling collapse is
very good.}
\label{fig:ncov_coll}
\end{figure}

The average number of distinct nodes visited by a random walker,
denoted $\Nc$, was first studied by Dvoretzky and Erd{\"o}s
\cite{dvoretzky} for a random walker on an infinite $d$-dimensional
regular lattice.  They found that $\Nc \sim \sqrt{t}$ in 1-$d$, $\Nc
\sim t/ \ln t$ in 2-$d$ and $\Nc \sim t$ for $d>2$, in the limit of
$t\rightarrow \infty$.  How does the small-world network modify this
result?  Following the arguments of \cite{jasch01}, we expect that for
short times, $t \ll \xi^2$ (with $\xi=1/p$), the random walker will be
probing regions of the small-world network that are essentially
without shortcuts (linear regions).  Hence, the behavior should be
similar to the $p=0$ case where the walker covers $\Nc \sim \sqrt{t}$
sites. For long times, $t \gg L \xi$, we expect all the sites in the
network to be covered and $\Nc = L$.  For intermediate times, $ \xi^2
\ll t \ll L \xi$, the walker spends on the average $\xi^2$ time steps
per linear region, before a {\em new} region is accessed -- the
shortcuts act like a kind of branching process for the random walk.
As a consequence, in this time regime there should be $t/\xi^2$
segments covered and $\Nc \sim t$ \cite{jasch01}.  Combining these
regimes, we obtain the scaling form
\begin{equation}
   \Nc = L ~S(t/\xi^2;x), 
\end{equation}
where
\begin{equation}
S(y;x) ~\sim~ \left\{ \begin{array}{rc}
		\sqrt{y}/x, & y \ll 1,\\
	 	y/x       , & 1 \ll y \ll x, \\
                1         , & y \gg x,
		 \end{array} \label{eq:ncov}
         \right.
\end{equation}
and $y = t / \xi^2$.  This is the expected finite-size scaling form
for the average number of distinct visited nodes in the small-world
model in the limit $x > 1$.

In Fig.\ \ref{fig:ncov_coll}, we show a plot of our calculated $\Nc$
for $x=10$ (circles), $10^2$ (squares) and $10^3$ (stars). For each
value of $x$, we have used different values of $p$, ranging from
$p=0.0002$ to $p=0.01$, two for each $x$-value. The scaling collapse
is excellent.  These results suggest that the scaling properties of
the small-world network determine the scaling of the random walk;
hence, to find a scaling collapse, one must keep the average number of
shortcuts, $x$, constant.  This Figure also shows that, as $x$
increases, $\Nc$ deviates more and more from the $p=0$ result.

\begin{figure}[t]
\centerline{\includegraphics[height=6.5cm]{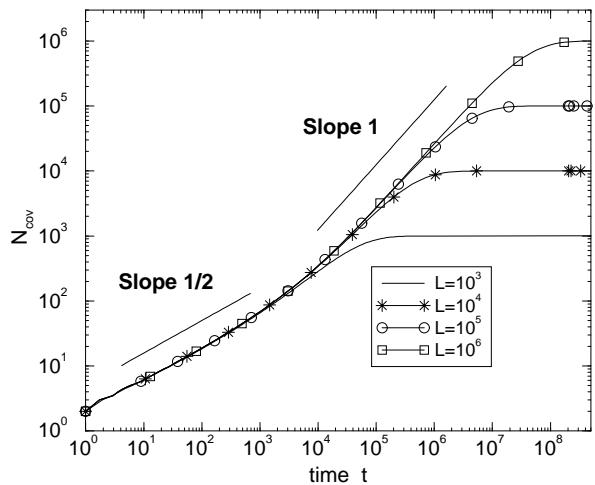}}
\caption{Effects of finite size on the average number of covered
sites, $\Nc$, which is plotted as a function of time $t$ at fixed
density of shortcuts $p=0.01$, for four different system sizes:
$L=10^3$, $10^4$, $10^5$ and $10^6$.  The saturation seen at all four
values of $L$ is a finite-size effect.}
\label{fig:ncov_p}
\end{figure}

If, instead, we hold the density of shortcuts constant and vary the
system size, we can explore the finite-size effects on the average
number of covered sites, $\Nc$.  In Fig.\ \ref{fig:ncov_p} we plot the
succession of curves $L=10^3$, $10^4$, $10^5$ and $10^6$, all with
$p=0.01$. Upon examination, we find that the slope of the resulting
scaling curve changes from $1/2$ to $1$.  That is, for small $t$
($t\ll \xi^2$), $\Nc \sim \sqrt{t}$, after which, at larger $t$, there
is a crossover to $\Nc \sim t$, and finally, at even larger $t$,
finite-size effects become apparent.  This behavior observed in the
simulations agrees very well with the arguments preceding Eq.\
(\ref{eq:ncov}).  Note that, when $\xi \lesssim L$, we do not see a
crossover to the linear regime in which $\Nc \sim t$ before finite
size effects start to become dominant.

We now derive an approximate expression for $\Nc$ which is consistent
with Eq.\ (\ref{eq:ncov}) and also in good agreement with the
large-$x$ simulation results.  We note that, on average, there are
$2x$ shortcut ends in the network, defining $2x$ contiguous
``regions'' in the 1-$d$ network.  If a walker takes a shortcut (or
``jump''), it transports the walker from one region to another,
randomly chosen, region in the network.  When the number of jumps
$n_j$ is small compared to $2x$, there is a high probability that the
regions visited by the walker are ``distinct'' (as defined more
precisely below).  However, when $n_j \sim 2x$, there is a high
probability that some regions will be visited more than once.  These
multiple visits lead to saturation effects since the
already-visited sites do not contribute to $\Nc$.

\begin{figure}[t]
\centerline{\includegraphics[height=6.5cm]{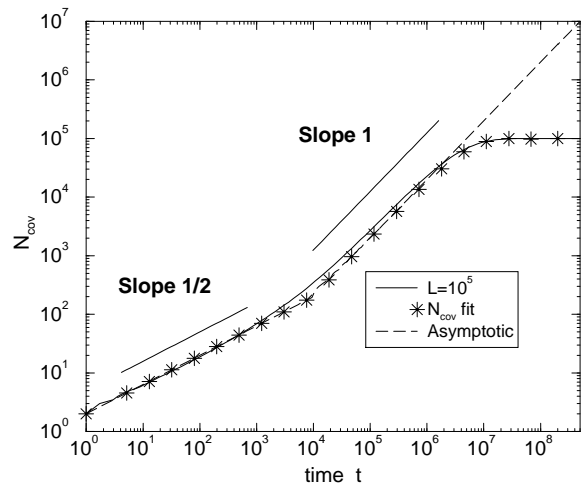}}
\caption{Comparison of the approximation (stars) to $\Nc$,
Eq. (\ref{eq:hp1}), with numerical data (solid line) for $L=10^5$ and
$p=0.01$. The asymptotic $(L \rightarrow \infty)$ curve (dotted line)
satisfies $\Nc \propto \sqrt{t}$ for small $t$ ($t \ll \xi^2 = 1/p^2$)
and $\Nc \propto t$ for large $t$ ($t \gg \xi^2$). The change of slope
between the two behaviors occurs near $t \sim \xi^2$.}
\label{fig:ncov_p2}
\end{figure}

We now derive an expression for the mean number of distinct regions
visited by the walker in $n_j$ jumps.  Because each jump transports
the walker from one region to another, randomly chosen region, the
number of distinct regions is obtained by solving the following
problem: If $n_j$ integers are chosen independently and randomly from
the set $\{1,2,..., 2x\}$, what is the mean number of {\em distinct}
integers chosen?  This latter problem is easily solved: The
probability that a given integer is {\em not} chosen in a given trial
is $q = 1 - \frac{1}{2x}$; hence, the probability that a given integer
is not chosen in $n_j$ trials is $ q^{n_{j}} \approx e^{- n_{j} /
(2x)}$ (where we have assumed $x \gg 1$).  Thus, the mean number of
distinct integers chosen after $n_j$ trials is
\begin{equation}
	I_{\mbox{cov}} ~=~ 2x~\left(  1  - e^{ - \frac{n_{j}}{2x}} \right).
\end{equation}
For the original problem, the above expression gives us the number of
distinct regions visited by the walker after $n_{j}$ jumps, and each
distinct region visited by the walker corresponds to covering $L
/(2x)$ sites.  Furthermore, after time $t$, the random walker has, on
average, taken $n_{j} = \lfloor 4 p^{2}t \rfloor$ jumps.  Hence, the
probability that the walker visits a new region after the
$n_{j}^{\mbox{th}}$ jump is $\exp(- \lfloor 2y \rfloor / x)$.  Also,
the time spent by the walker in this new region is $t' = t -
\frac{1}{4 p^2} \lfloor 4 p^2 t\rfloor$, while the number of sites
covered in this time interval is given by $\sqrt{t'}$, {\it i.e.} is
proportional to the expression for the $p=0$ case.  Combining these
estimates, we find that the mean number of sites visited after time
$t$ is given by
\begin{equation}
	S(y,x) ~\approx~ 1 - e^{-\frac{2}{x}\lfloor y \rfloor} + 
	       \frac{2}{x} \sqrt{ y - \lfloor y \rfloor } 
		~e^{-\frac{2}{x}\lfloor y \rfloor}  \label{eq:hp1}
\end{equation}

Expression (\ref{eq:hp1}) is clearly only an approximation to the
exact function $S(y,x)$.  However, it does captures the key processes
leading to the growth and saturation of $\Nc$ and hence is a useful
approximation.  In Fig.\ \ref{fig:ncov_p2}, we compare this function
to the numerically obtained curves for $S(y,x)$ at $p=0.01$, taking $L
= 10^5$ and $L \rightarrow \infty$.  The agreement is quite good and
shows that Eq.\ (\ref{eq:hp1}) is a reasonable approximation to the
scaling function.

\subsection{Mean-Square Displacement}

\begin{figure}[tb]
\centerline{\includegraphics[height=6.5cm]{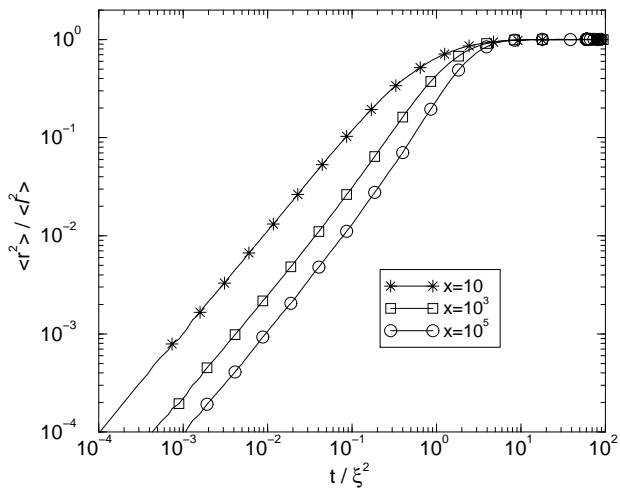}}
\caption{Mean-square displacement, $\la r^2(t) \ra$, for a random
walker on an SWN for several values of $p \le 0.01$ and the three
choices for average numbers of shortcuts: $x=p L = 10$, $10^3$ and
$10^5$.  Note that each constant-$x$ curve consists of two distinct
$(p,L)$ combinations. $\la r^2(t)\ra$ saturates at large $t$ because
of finite-size effects.}
\label{fig:RW_x}
\end{figure}

We now compute the mean-square displacement, $\la r^2(t) \ra$, of the
random walker as a function of time $t$.  To calculate this quantity,
we first, at each time step, find the {\em minimal} distance from the
current position of the random walker to the origin ({\it i.e.} the
smallest number of steps needed for the random walker to reach the
origin) using a breadth-first search method.  Then we allow the walker
to move through the network until $\la r^2(t) \ra$ has saturated.
Finally, we average over different initial positions of the walker and
realizations of the network.

Now, we know that for a random walk on an infinite, hypercubic,
$d$-dimensional lattice $\la r^2(t) \ra ~=~ (1/2)~D~ t$, as can be
shown using {\it e.g.} a generating function formalism
\cite{barber70,montroll64}.  However, on a {\em finite} lattice, $\la
r^2(t) \ra$, must approach a constant for large $t$.  In this limit,
each node of the network has equal probability of being occupied by
the random walker.  This insight immediately gives $\la r^2 \ra =
\lbsq$, where $\lbsq$ is the squared minimum distance between a pair
of nodes, averaged over all possible pairs and network realizations.

On an SWN, the other relevant length scale for the random walker is
$\xi = 1/p$, the {\em average} distance the walker travels to reach a
shortcut.  These two lengths suggest the following scaling Ansatz:
$\la r^2 \ra ~=~ \lbsq~ R (t / \xi^2 ; p L)$.  We can also infer the
behavior of $R(y, x)$ using simple arguments.  For times $t \ll
\xi^2$, the walker is exploring regions of the network without
shortcuts, and we expect diffusive behavior similar to that of a
regular network, giving $R(y,x) \sim y/\lbsq$.  When $t \sim
\sqrt{\lbsq}/p$, we expect the mean-square displacement to saturate,
and $R(y,x) = 1$.  The transition between the two types of behavior
is not sharp, since the walker may reach a shortcut before it has
travelled a distance $\xi$.

We have numerically confirmed this scaling collapse for a wide range
of $x$ values.  In Fig.\ \ref{fig:RW_x}, we plot $R (t /\xi^2; x)$ for
a sequence of networks with $x = p L = 10$, $10^3$ and $10^5$. For
each constant-$x$ curve, we show use two distinct values of $p \le
0.01$.  We see that initially, the walker probes a regular network,
and $R (t /\xi^2; x)$ is linear in $t$.  When $t/\xi^2 \sim 1$, the
walker begins to reach some shortcuts in the network, and there is a
crossover to superdiffusive behavior.  Finally, at still larger
$t/\xi^2$, finite-size effects become important and $R (t /\xi^2; x)$
saturates.

\begin{figure}[tb]
\centerline{\includegraphics[height=6.5cm]{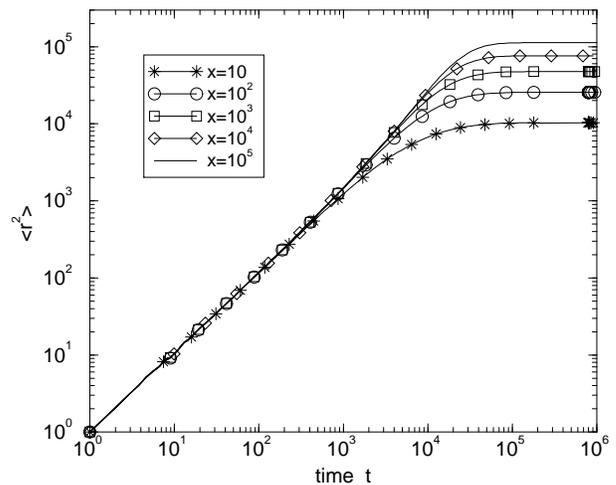}}
\caption{Mean-square displacement, $\la r^2(t) \ra$, as a function of
time $t$ for a random walker on a succession of networks, all with the
same density of shortcuts ($p=0.01$), and system sizes varying from
$L=10^3$ to $L=10^7$.  Except for very large $t$, all curves collapse
onto a single scaling curve, the curve which would be obtained for
$L\rightarrow\infty$, but at sufficiently large $t$, finite-size
effects become important.}
\label{fig:RW_p}
\end{figure}

To further explore finite-size effects on the mean-square
displacement, we have studied $\la r^2(t)\ra$ in a succession of
networks, each with the same density of shortcuts, $p=0.01$, but with
different linear size $L$.  In Fig. \ref{fig:RW_p}, we plot the
calculated $\la r^2(t) \ra$ for values of $L$ differing by factors of
$10$ and ranging from $10^3$ to $10^7$. The resulting curves are all
very similar to the $L\rightarrow\infty$ curve, until finite size
effects produce saturation with $\la r^2 \ra = \lbsq$.  Only for the
largest $x$-values are we able to reach the superdiffusive regime, and
even for $x=10^5$, this regime is so narrow that we cannot with
confidence determine the exponent of the expected power-law time
dependence.

\begin{figure}[t]
\centerline{\includegraphics[height=6.5cm]{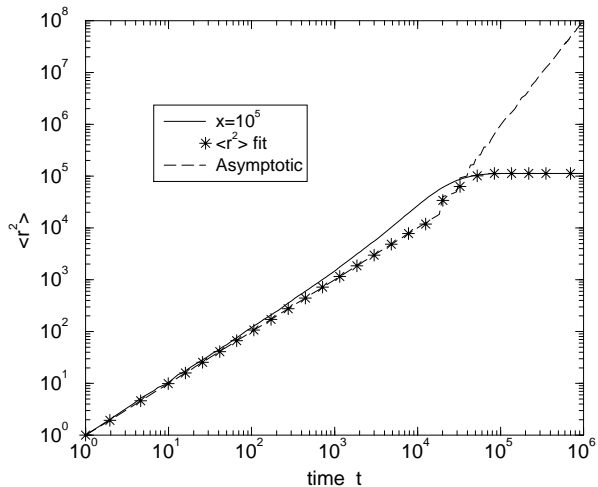}}
\caption{Comparison of the approximation (stars) to $\la r^2(t)\ra$,
Eq. (\ref{eq:r}), with numerical data (solid line) for $L=10^7$ and
$p=0.01$. The asymptotic $(L \rightarrow \infty)$ curve (dotted line)
satisfies $\la r^2(t)\ra \propto t$ for small $t$ ($t \ll \xi^2$)
and $\la r^2(t)\ra \propto t^2$ for large $t$ ($t \gg \xi^2$).}
\label{fig:RW_p2}
\end{figure}

The development of an approximate analytical expression for $\la
r^2(t) \ra$ is difficult, since we need the minimal distance between
two lattice points. However, we know the limiting forms for $\la
r^2(t) \ra$. Further, by making the approximation that the random
walker only uses a short-cut once, we can extend the arguments used to
derive the approximate expression for $\Nc$,
Eq. (\ref{eq:hp1}). Hence, we can write down the following ansatz for
$\la r^2(t)\ra$:
\begin{equation} 
 R(y,x) ~=~ 1 - e^{-\frac{1}{p^2 \lbsq} \lfloor y \rfloor^2 } \left(1
  - \frac{1}{p^2 \lbsq} \left( y - \lfloor y \rfloor \right) \right).
\label{eq:r}
\end{equation}
In Fig.\ \ref{fig:RW_p2}, we compare this ansatz with the simulations
for $L = 10^7$; evidently it agrees reasonably well with the numerical
results.

We can write down the scaling function in several limiting regimes,
without necessarily using the above ansatz for the specific functional
form.  For example, we must have
\begin{equation}
	R(y;x) \sim y/(p^2\lbsq), ~~~y \ll 1,
\end{equation}
and 
\begin{equation}
	R(y;x) = 1, ~~~~y \gg p\sqrt{\lbsq}.
\end{equation}
However, the known limiting behaviors cannot give the scaling function
in the intermediate regime.  We therefore use the above ansatz to
propose
\begin{equation}
	R(y;x) \sim y^2/(p^2\lbsq), ~~~1 \ll y \ll p\sqrt{\lbsq}.
\end{equation}
We are not able to verify this last behavior numerically from our
present results.

\subsection{First Return Time Distribution}

\begin{figure}[t]
\centerline{\includegraphics[height=6.5cm]{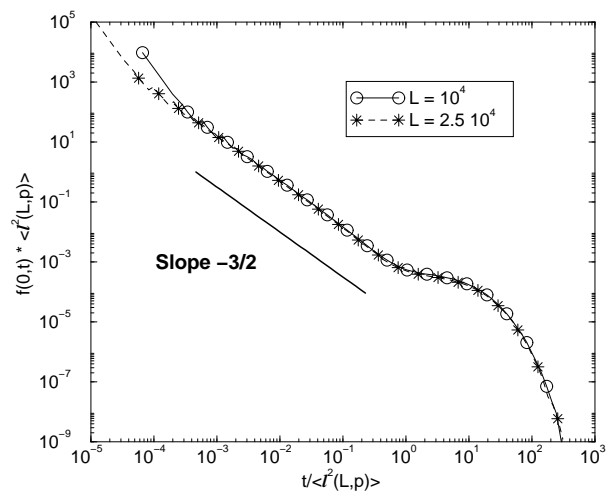}}
\caption{Scaling collapse of the first-return time distribution,
$f(0,t)$ plotted against the dimensionless time $u=t/\lbsq(L,p)$ for
networks with the average number of shortcuts $x = p L = 100$ using
$L= 10^4$ and $L=2.5\times 10^4$ nodes.}
\label{fig:return_x}
\end{figure}

Next, we turn to another property of a walker on an SWN, namely the
distribution of first passage (or first return) times.  This is the
probability that a walker will return to a given site $m$ for the {\em
first time} a time $t$ after leaving that site.  We denote this
distribution $f(m,t;L,p)$. Note, that this distribution does not
saturate to a finite value when $t\rightarrow\infty$, and hence, it
shows that the scaling collapse of random walk properties is not only
limited to those which saturate to a finite value. In order to find a
smooth distribution, we must make a slight change to the random walker
rules: For each time-step, the walker is now allowed to stay on the
current node with probability $1/(k(m)+1)$.

\begin{figure}[t]
\centerline{\includegraphics[height=6.5cm]{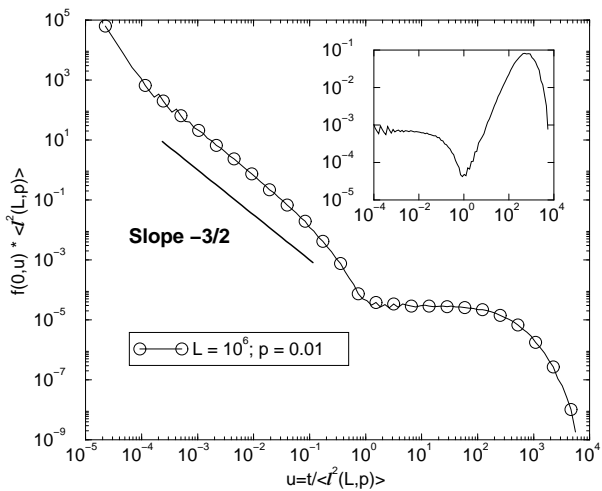}}
\caption{First-return time distribution, $f(0,t)$, plotted against the
dimensionless time $u=t/\lbsq(L,p)$. For waiting times $t \ll
\lbsq(L,p)$, the first-return time distribution $f(0,t)$ behaves like
that of a $p=0$ network, with $f(0,t) \sim t^{-3/2}$.  For
intermediate waiting times ($\lbsq(L,p) \ll t \ll L\xi$), $f(0,t)$
is independent of $t$, since each node is occupied with equal
probability.  For $t > L\xi$, lattice finite-size effects
dominate. The inset shows $(t/\lbsq)^{3/2} f(0,t/\lbsq)$.}
\label{fig:first}
\end{figure}

An extensive review of first passage processes has been given by
Redner \cite{redner}.  Unlike the other quantities presented in this
article, the first-return time does not saturate to a finite value,
instead it approaches zero. However, the scaling collapse is still
present.  This collapse is demonstrated in Fig.\ \ref{fig:return_x},
where we plot $f(0,t)$ for two different $(p,L)$ combinations, holding
$x$ constant.

The ``small-world'' effect on the first-return time distribution is
most clearly seen at large values of $x$.  In Fig.\ \ref{fig:first},
we plot $f(0,t)$ for a network of $L=10^6$ sites and a shortcut
density of $p=0.01$.  For short waiting times, as expected, $f(0,t)$
behaves like that of a $p=0$ network (no shortcuts).  However, for
intermediate waiting times such that $\lbsq \ll t \ll L\xi$, we find
that $f(0,t)$ is {\em independent} of $t$. This independence indicates
that the position of the random walker is completely randomized: the
memory of the walker's starting position is no longer retained in the
system and the walker is equally likely to occupy any site in the
network. This occurs when $t \sim \lbsq$, as discussed above for $\la
r^2(t) \ra$. Also, $L\xi$ is the characteristic time for the
asymptotic decay of $f(m,t;L,p)$, since it is the saturation timescale
for $\Nc$.  Note that the separation of $\lbsq(L,p)$ and $L\xi$
increases with increasing number of shortcuts $x$ in the system,
making the ``knee'' of the first-return distribution more
pronounced.

For a random walk on a finite interval $[0,L]$ with an absorbing wall
at zero and a reflecting wall at $L$ there is a similar effect in the
first return time\cite{redner,redner03}; at $t \sim L^2$ there is an
enhancement of the first return probability. In our case, the origin
of this effect is the splitting of the timescales $\lbsq$ and $L\xi$,
while for the walk on the finite interval the cause for this effect
are the contributions from reflected trajectories.

\section{Summary}

In summary, we have studied the behavior of a random walker on a
small-world network, using a combination of numerical methods and
scaling assumptions.  We conjecture that the scaling law of Eq.\
(\ref{eq:genscale}) is obeyed by measurable, saturating properties
${\cal O}(p, L, t)$ of a random walk on an SWN.  Among these
properties are the average number of distinct visited sites, $\Nc$,
and the mean-square displacement $\la r^2 \ra$ for random walks on
SWN's, both of which we have studied over a wide range of node numbers
$L$ and shortcut densities $p$. In both cases, we find that
Eq. (\ref{eq:genscale}) is satisfied, and the quantities depend only
on the single variable $x=p L$. Additionally, we find that
non-saturating properties also show a scaling collapse, as exemplified
by the first-return time.

Thus, we have shown that the dynamical behavior of a random walker on
an SWN has the same scaling behavior as that exhibited by purely
geometrical properties of the network (as described in {\it e.g.}
Ref.\ \cite{kulkarni}).  This scaling behavior should be useful in
interpreting a variety of other properties on SWN's, and may be of
value in studying real-world phenomena for which an SWN is a good
model.

\acknowledgments

This work has been supported by NSF through Grant No. DMR01-04987
(E.A. and D.S.) and the U.S. Department of Energy, Office of Science,
Division of Materials Research (R.V.K).

\end{document}